# CrowdER: Crowdsourcing Entity Resolution

Jiannan Wang [†#]     Tim Kraska [†]     Michael J. Franklin [†]     Jianhua Feng [#]

[†]*AMPLab, UC Berkeley*     [#]*Department of Computer Science, Tsinghua University*

wjn08@mails.tsinghua.edu.cn; {kraska, franklin}@cs.berkeley.edu; fengjh@tsinghua.edu.cn

## ABSTRACT

Entity resolution is central to data integration and data cleaning. Algorithmic approaches have been improving in quality, but remain far from perfect. Crowdsourcing platforms offer a more accurate but expensive (and slow) way to bring human insight into the process. Previous work has proposed batching verification tasks for presentation to human workers but even with batching, a human-only approach is infeasible for data sets of even moderate size, due to the large numbers of matches to be tested. Instead, we propose a hybrid human-machine approach in which machines are used to do an initial, coarse pass over all the data, and people are used to verify only the most likely matching pairs. We show that for such a hybrid system, generating the minimum number of verification tasks of a given size is NP-Hard, but we develop a novel two-tiered heuristic approach for creating batched tasks. We describe this method, and present the results of extensive experiments on real data sets using a popular crowdsourcing platform. The experiments show that our hybrid approach achieves both good efficiency and high accuracy compared to machine-only or human-only alternatives.

## 1. INTRODUCTION

Entity resolution (also known as entity reconciliation, duplicate detection, record linkage and merge/purge) in database systems is the task of finding different records that refer to the same entity. Entity resolution is particularly important when cleaning data or when integrating data from multiple sources. In such scenarios, it is not uncommon for records that are not exactly identical to refer to the same real-world entity. For example, consider the table of product data shown in Table 1. Records $r_1$ and $r_2$ in the table have different text in the Product Name field, but refer to the same product. Our goal is to find all such duplicate records.

There has been significant work in developing automated algorithms for entity resolution (see [11] for a recent survey). A basic machine-based technique is to compute a pre-defined similarity metric, such as Jaccard similarity, for each pair of records [2,5,26]. Records whose similarity values are above a specified threshold are considered to refer to the same entity. More sophisticated techniques use machine learning. For example, some approaches model entity resolution as a classification problem, training a classifier to distinguish between "duplicate" or "non-duplicate" pairs [4,6]. Despite all of this progress, machine-based techniques remain far from perfect. For example, a recent study [18] describes the difficulty that state-of-the-art techniques have in many domains such as identifying duplicate products based on their textual descriptions.

Table 1: A table of products.

| ID | Product Name | Price |
|---|---|---|
| $r_1$ | iPad Two 16GB WiFi White | $490 |
| $r_2$ | iPad 2nd generation 16GB WiFi White | $469 |
| $r_3$ | iPhone 4th generation White 16GB | $545 |
| $r_4$ | Apple iPhone 4 16GB White | $520 |
| $r_5$ | Apple iPhone 3rd generation Black 16GB | $375 |
| $r_6$ | iPhone 4 32GB White | $599 |
| $r_7$ | Apple iPad2 16GB WiFi White | $499 |
| $r_8$ | Apple iPod shuffle 2GB Blue | $49 |
| $r_9$ | Apple iPod shuffle USB Cable | $19 |

The limitations of machine-based approaches combined with the availability of easily-accessible crowdsourcing platforms have caused many to turn to human-based approaches. Indeed, de-duplication (of addresses, names, product descriptions, etc.) is an important use case for popular crowdsourcing platforms such as Amazon Mechanical Turk (AMT) and Crowdflower. Such platforms support crowdsourced execution of "microtasks" or Human Intelligence Tasks (HITs), where people do simple jobs requiring little or no domain expertise, and get paid on a per-job basis. Entity resolution is easily expressed as a query in a crowd-enabled query processing system such as CrowdDB [13] or Qurk [20]. For example, in CrowdDB's CrowdSQL language, the following self-join query identifies duplicate product records.

```
SELECT p.id, q.id FROM product p, product q
WHERE p.product_name ~= q.product_name;
```

Note that ~= is an operator that can ask the crowd to decide whether or not p.product_name and q.product_name refer to the same product.

A naive way to process such a query is to create a HIT for each pair of records, and for each HIT, to ask people in the crowd to decide whether or not the two records both refer to the same entity. For a table with $n$ records, the naive execution method will lead to $\mathcal{O}(n^2)$ HITs. In a recently published paper [19], Marcus et al. proposed two batching strategies to reduce the number of HITs for matching operations (they focused on `join` operations). The first is to simply place multiple record pairs into a single HIT. To perform such a pair-based HIT, a worker must check each pair in the HIT individually. Suppose a pair-based HIT consists of $k$ pairs of records. The number of HITs will be reduced to $\mathcal{O}(n^2/k)$. Their second batching approach also placed multiple pairs of records in a single HIT but they asked workers to find all matches among all of the records. In this latter approach, if a HIT consists of $k$ records the number of HITs will be reduced to $\mathcal{O}(n^2/k^2)$. Their results indicated that batching could provide significant benefits in cost with only minimal negative impact on accuracy. However, their approach suffers from a scalability problem. Even with a modest database size of 10,000 records, assuming a reasonable HIT size $k = 20$, their approaches would require 5,000,000 and 250,000 HITs respectively. At even \$0.01 per HIT, this query would cost \$50,000 or \$2,500 to execute, if in fact, it could even be successfully executed on existing platforms.

While the insight that batching can reduce the number of HITs is a correct one, clearly batching on its own is not sufficient to enable Entity Resolution to be done at scale. Instead, what is needed is a *hybrid human-machine* approach, that uses machine-based techniques to weed out obvious non-duplicates, while using precious human resources to examine just those cases where human insight is needed. Similar hybrid approaches have been shown to be effective for problems such as image search [27] and language translation [23]. Following this line of research, we propose a hybrid human-machine Entity Resolution approach called CrowdER. CrowdER first uses machine-based techniques to discard those pairs of records that look very dissimilar, and only asks the crowd to verify the remaining pairs.

Having machine-based similarity estimates raises a further opportunity. Namely, the similarity estimates can be used to further optimize the grouping of specific records into HITs. Similar to Marcus et al., we consider batching of records into HITs as groups of individual pairs ("pair-based HIT generation") or in a format that requires workers to find matches among all the records ("cluster-based HIT generation") but with the major difference that we use machine-generated similarity estimates to drive the creation of the HITs. We formulate the HIT generation problem and show that generating the minimum number of cluster-based HITs is NP-Hard. Thus, we develop a heuristic-based algorithm for generating cluster-based HITs and evaluate the algorithm analytically and through an extensive experimental analysis using real data sets and a popular crowdsourcing platform.

The main contributions are the following:

- We propose a hybrid human-machine system for entity resolution by combining human-based techniques and machine-based techniques.

- We formulate the cluster-based HIT generation problem, prove that it is an NP-Hard problem and develop a two-tiered heuristics-based solution for it.

- We compare pair-based HIT generation and cluster-based HIT generation analytically and experimentally.

- We have implemented our approaches and compared them with the state-of-the-art techniques on real datasets using the AMT platform. Experimental results show that our approach reduces cost while providing good answer quality. In particular, our hybrid human-machine approach makes it practical to bring humans into the Entity Resolution process.

The remainder of this paper is organized as follows. We propose a hybrid human-machine approach for entity resolution in Section 2. Section 3 introduces the HIT generation problem and proves that it is NP-Hard. We present an approximation algorithm in Section 4 and devise a more practical, two-tiered approach in Section 5. Section 6 compares pair-based HIT and cluster-based HIT generation analytically. We describe the results of our experimental studies in Section 7. Related work is reviewed in Section 8 and we present conclusions and future work in Section 9.

## 2. ENTITY RESOLUTION TECHNIQUES

In this section, we first review existing machine-based techniques for entity resolution and then describe a hybrid workflow combining people and machines.

### 2.1 Machine-based Techniques

Machine-based Entity Resolution techniques can be broadly divided into two categories, similarity-based and learning-based.

#### 2.1.1 Similarity-based

Similarity-based techniques require a similarity function and a threshold. The similarity function takes a pair of records as input, and outputs a similarity value. The more similar the two records, the higher the output value. The basic approach is to compute the similarity of all pairs of records. If a pair of records has a similarity value no smaller than the specified threshold, then they are considered to refer to the same entity.

For example, in Table 1, suppose that the similarity of two records is specified as Jaccard similarity between their Product Names, and the specified threshold is 0.5. Jaccard similarity over two sets is defined as the size of the set intersection divided by the size of the set union. For example, the Jaccard similarity between the Product Names of $r_1$ and $r_2$ is

$$\mathrm{J}(r_1, r_2) = \frac{|\{\texttt{iPad, 16GB, WiFi, White}\}|}{|\{\texttt{iPad, 16GB, WiFi, White, Two, 2nd, generation}\}|} = 0.57.$$

The similarity-based technique will consider $\langle r_1, r_2 \rangle$ as referring to the same entity since their Jaccard similarity is no smaller than the threshold, i.e., $\mathrm{J}(r_1, r_2) \geq 0.5$. Similarly, $\langle r_1, r_3 \rangle$ will not be considered a match since $\mathrm{J}(r_1, r_3) = 0.25 < 0.5$.

Since it is expensive to compute the similarity for every pair of records, research on similarity-based techniques [2, 5, 26] mainly focuses on how to reduce the number of pairs evaluated.

#### 2.1.2 Learning-based

Learning-based techniques model entity resolution as a classification problem [4,6]. They represent a pair of records as a feature vector in which each dimension is a similarity value of the records on some attribute. If we choose $n$ similarity functions on $m$ attributes, then the feature vector will be a $nm$-dimensional feature vector. For example, for the records in Table 1, suppose we only choose Jaccard

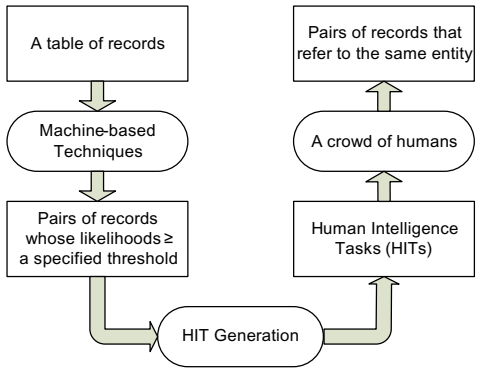

Figure 1: Hybrid human-machine workflow.

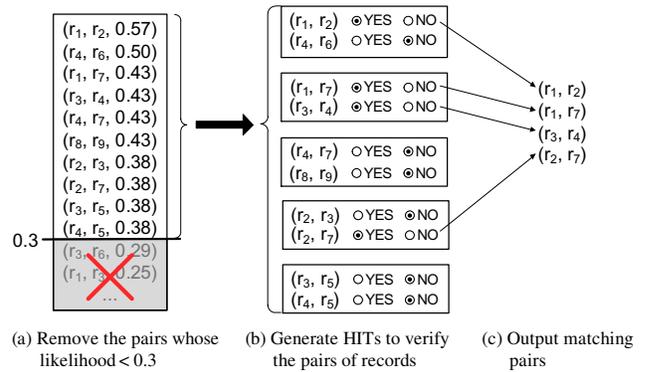

(a) Remove the pairs whose likelihood < 0.3    (b) Generate HITs to verify the pairs of records    (c) Output matching pairs

Figure 2: An example of using the hybrid human-machine workflow to find duplicate pairs in Table 1.

similarity on `Product Name`. Then each pair of records will be represented as a feature vector that contains only a single dimension. Learning-based techniques require a training set to train the classifier. The training set consists of positive feature vectors and negative feature vectors indicating matching pairs and non-matching pairs respectively. The trained classifier can then be applied to label new record pairs as matching or non-matching.

## 2.2 Hybrid Human-Machine Workflow

As described in the introduction, people are often better than algorithms at detecting when different terms actually refer to the same entity. However, compared to algorithmic techniques people are much slower and more expensive. A hybrid human-machine approach has the potential to combine the efficiency of machine-based approaches with the answer quality that can be obtained from people. The intuitive idea is that among $\frac{n \cdot (n-1)}{2}$ pairs of records that can be generated from a set of $n$ records, a large number of pairs are very dissimilar. Such pairs can be easily pruned using a machine-based technique. People can then be brought in to examine the remaining pairs.

Based on this idea, we propose a human-machine workflow as shown in Figure 1. The workflow first uses machine-based techniques to compute for each pair the likelihood that they refer to the same entity.[1] For example, the likelihood could be the similarity value given by a similarity-based technique. Then, only those pairs whose likelihood exceeds a specified threshold are sent to the crowd. In the experimental section, we show that by specifying a relatively low threshold we can dramatically reduce the number of pairs that need to be verified with only a minor loss of quality. Given the set of pairs to be sent to the crowd, the next step is to generate HITs so that people can check them for matches. *HIT Generation* is a key component of our workflow. Finally, generated HITs are sent to the crowd for processing and the answers are collected.

EXAMPLE 1. *Consider the nine records in Table 1. Instead of asking people to check $\frac{9*8}{2} = 36$ pairs, our workflow first employs a machine-based technique to compute the likelihood of each pair of records. Here we use the `Jaccard` similarity between `Product Names` of a pair of records as their likelihood. Then the workflow prunes the pairs whose likelihood is smaller than the specified threshold. Suppose the threshold is 0.3. Figure 2(a) shows the remaining ten pairs*

---

[1] In practice, we can adopt some indexing techniques such as blocking and Q-gram based indexing [7] to avoid all-pairs comparison.

*of records. That is, the workflow only needs to generate HITs to verify the ten pairs (rather than 36). In this example, we batch two pairs into each HIT, and generate five HITs as shown in Figure 2(b). As an example, in the first HIT, the crowd selects "YES" for $(r_1, r_2)$ and "NO" for $(r_4, r_6)$ which indicates that $r_1$ and $r_2$ are the same entity while $r_4$ and $r_6$ are not. After the crowd finishes all the HITs, the four matching pairs (as determined by the crowd) are returned (Figure 2(c)).*

## 3. HIT GENERATION

Recall that a key step in the human-machine workflow is HIT Generation. That is, given a set of pairs of records, they must be combined into HITs for crowdsourcing. In this section we discuss the generation problem for two types of HITs, pair-based HITs and cluster-based HITs.

### 3.1 Pair-based HIT Generation

A pair-based HIT consists of multiple pairs of records to be compared batched in to a single HIT. For each pair of records, the crowd needs to verify whether they refer to the same entity or not. Figure 3 shows an example of the user interface we generate for a pair-based HIT on AMT. At the top, there is a brief description of the HIT. More detailed instructions can be displayed by clicking "Show Instructions". The HIT shown consists of two pairs of records. For each pair, the worker needs to choose either "They are the same product" or "They are different products". The HIT can be submitted only if a selection has been made for all pairs of records in the HIT. Note that in Figure 3, the second pair of records has not been verified, so the submit button is disabled, and its caption shows "1 left". We also recommend (but do not require) that workers provide reasons for their choices.

Generating pair-based HITs is straightforward. Suppose a pair-based HIT can contain at most $k$ pairs. Given a set of pairs, $\mathcal{P}$, we need to generate $\lceil \frac{|\mathcal{P}|}{k} \rceil$ pair-based HITs. For example, for the ten pairs of records with above-threshold likelihood in Figure 2(a), if $k = 2$, we would need to generate five pair-based HITs.

### 3.2 Cluster-based HIT Generation

A cluster-based HIT consists of a group of individual records rather than pairs. Workers are asked to find all duplicate records in the group. Figure 4 shows the user interface we generate for a cluster-based HIT on AMT. As with pair-based HITs, there is a brief description at the top, and more detailed instructions are available. The example HIT contains four records. A drop-down list at the front of each record allows a worker to assign each record a label. Initially, all records are unlabeled. When a label is selected for a record, the background color of the record is changed to the corresponding color for that label. Workers indicate duplicate records by assigning them the same label (and thus, the same color).

Figure 3: A pair-based HIT with two record pairs.

To make the labeling process more efficient, our interface supports two additional features, (1) sorting records by column values by clicking a column header; (2) moving a record by dragging and dropping it. The first feature can be used for example, to sort the records based on a specific attribute such as product price. The second feature can be used, for example, to place the records that share a common word, e.g. "ipad" near each other for easier comparison.

Next we study how to generate cluster-based HITs. A cluster-based HIT allows a pair of records to be matched iff both records are in the HIT. In a crowdsourced system like AMT, payment is made for each successfully completed HIT. Thus, there is a financial incentive to minimize the number of HITs. However, placing too many records in a cluster-based HIT makes it difficult for workers to complete, resulting in higher-latencies and lower quality answers. Thus, we bound the number of records placed in a cluster-based HIT. We can then formulate the HIT generation problem as follows:

DEFINITION 1 (CLUSTER-BASED HIT GENERATION).
*Given a set of pairs of records, $\mathcal{P}$, and a cluster-size threshold, $k$, the cluster-based HIT generation problem is to generate the minimum number of cluster-based HITs, $H_1, H_2, \cdots, H_h$, that satisfy two requirements: (1) $|H_\ell| \leq k$ for any $\ell \in [1, h]$, where $|H_\ell|$ denotes the number of records in $H_\ell$; (2) for any $(r_i, r_j) \in \mathcal{P}$, there exists $H_\ell$ ($\ell \in [1, h]$) s.t. $r_i \in H_\ell$ and $r_j \in H_\ell$.*

For example, consider the ten pairs of records in Figure 2(a). Given the cluster-size threshold $k = 4$, suppose we generate three cluster-based HITs, $H_1 = \{r_1, r_2, r_3, r_7\}$, $H_2 = \{r_3, r_4, r_5, r_6\}$ and $H_3 = \{r_4, r_7, r_8, r_9\}$. As their sizes are no larger than $k = 4$, the first requirement of Definition 1 holds. For any of the ten pairs, at least one of the three cluster-based HITs contain them, thus the second requirement of Definition 1 holds. Furthermore, it is impossible to find fewer cluster-based HITs that satisfy the two requirements. Therefore, based on Definition 1, $H_1$, $H_2$ and $H_3$ are the solution to the cluster-based HIT generation problem.

Figure 4: A cluster-based HIT with four records.

Unfortunately, the cluster-based HIT generation problem is NP-Hard. In the next section, we present an approximation algorithm for this problem.

THEOREM 1. *The cluster-based HIT generation problem is NP-Hard.*

PROOF. We prove it by reduction from the $k$-clique covering problem [15]. A $k$-clique is defined as a complete graph that contains $k$ vertices. We say that a $k$-clique covers an edge of a graph if the two vertices of the edge are both in the clique. Given a graph, the $k$-clique covering problem is to find the minimum number of $k$-cliques to cover all the edges of the graph. To reduce this problem to the cluster-based HIT generation problem, we take each vertex of the graph as a record, and construct a set of record pairs, $\mathcal{P}$, that consists of all edges of the graph. Let $H_1, H_2, \cdots, H_h$ be the solution to the reduced cluster-based HIT generation problem. Next we show that based on $H_1, H_2, \cdots, H_h$, we can obtain the solution to the original $k$-clique covering problem in polynomial time.

For each $H_\ell$ ($\ell \in [1, h]$), we generate a clique, $C_\ell$ that consists of the vertices corresponding to the records in $H_\ell$. Obviously, $C_1, C_2, \cdots, C_h$ are the minimum number of cliques that can cover all the edges of the graph. Since $|H_\ell| \leq k$, $C_\ell$ contains no larger than $k$ vertices. For each $C_\ell$ ($\ell \in [1, h]$), we can simply construct a $k$-clique, $C'_\ell$, by adding $k - |H_\ell|$ vertices into $C_\ell$, and finally obtain the solution to the $k$-clique covering problem, i.e. $C'_1, C'_2, \cdots, C'_h$. Therefore, the $k$-clique covering problem can be reduced to the cluster-based HIT generation problem in polynomial time. □

## 4. APPROXIMATION ALGORITHM

In this section, we first reduce our problem to the k-clique edge covering problem, and then apply its approximation algorithm to cluster-based HIT generation.

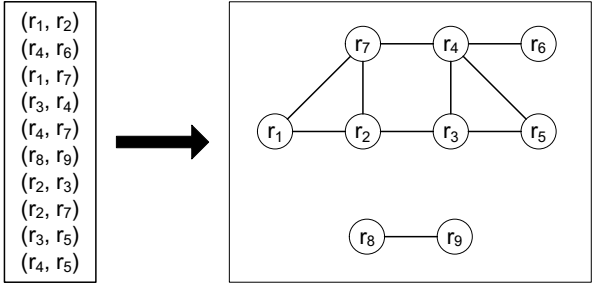

**Figure 5:** Build a graph based on the ten pairs of records to be verified in Figure 2(a).

In order to reduce our HIT generation problem to the $k$-clique covering problem, we build a graph based on the set of pairs that need to be verified. In the graph, each vertex represents a record, and each edge denotes a pair of records. A cluster-based HIT can be seen as a clique in the graph. We say the cluster-based HIT is able to check a pair if and only if the clique is able to cover the corresponding edge. (For simplicity, a cluster-based HIT is mentioned interchangeably with its corresponding clique in later text.) Therefore, the cluster-based HIT generation problem is reduced to finding the minimum number of cliques, whose sizes are no larger than $k$, to cover all edges of the graph. Note that we only need to consider the cliques whose sizes are equal to $k$ (i.e. $k$-cliques) since a larger clique can cover more edges than a smaller one. Therefore, the reduced problem is the same as the $k$-clique covering problem.

To solve the $k$-clique covering problem, we can model it as the set covering problem, and apply the corresponding approximation algorithm [8]. However, that algorithm is very expensive since it needs to generate $\binom{n}{k}$ covering cliques where $n$ is the number of the vertices in the graph. To address this problem, Goldschmidt ed al. [15] proposed an efficient ($\frac{k}{2} + \frac{k}{k-1}$)-approximation algorithm. The algorithm consists of two phrases.

**Phase 1:** The algorithm creates a sequence consisting of all the vertices and edges in the graph, denoted by $SEQ = \{e_1, e_2, \cdots, e_n\}$. Initially, $SEQ$ is empty. Then the algorithm iteratively selects a vertex, and adds the vertex and the edges that contain the vertex into $SEQ$, and removes them from the graph. The algorithm iterates until the graph has no vertices or edges.

**Phase 2:** Next, $SEQ$ is used to generate $k$-cliques to cover the edges of the graph. $SEQ$ has the useful property that for any subsequence with $k-1$ consecutive elements, i.e., $\{e_i, e_{i+1}, \cdots, e_{i+k-1}\}$, the edges in the subsequence contain at most $k$ different vertices [15]. Therefore, these edges can be covered by a $k$-clique. Based on this property, the algorithm divides $SEQ$ into $\lceil \frac{|SEQ|}{k-1} \rceil$ subsequences, where each subsequence has $k-1$ elements, and then finds a $k$-clique for each subsequence. Since all the edges of the graph are in $SEQ$, the algorithm can find $\lceil \frac{|SEQ|}{k-1} \rceil$ $k$-cliques to cover all the edges.

Example 2 shows how to use the above approximation algorithm to solve the cluster-based HIT generation problem.

EXAMPLE 2. *Consider the ten pairs in Figure 2(a). To generate cluster-based HITs for them, we first build a graph as shown in Figure 5. The graph contains ten edges, which represent the ten pairs. Next we create a sequence $SEQ$ which consists of all vertices and edges in the graph. Since the graph contains ten edges and nine vertices, there will be nineteen elements in $SEQ$. Suppose the cluster-size threshold is $k = 4$. We divide $SEQ$ into $\lceil \frac{|SEQ|}{k-1} \rceil = 7$ subsequences, where each subsequence (except the last one) contains $k-1 = 3$ elements. We generate a cluster-based HIT to cover the edges in each subsequence. Therefore, the approximation algorithm generates seven cluster-based HITs to verify the ten pairs in Figure 2(a).*

Note, however, that as described in Section 3.2 the optimal solution requires only three cluster-based HITs. In fact, as we show in our experimental evaluation (Section 7), this approximation algorithm generates many more cluster-based HITs than even a naive algorithm on the data sets we tested. Thus, in the following section, we propose a new cluster-based HIT generation algorithm.

## 5. A TWO-TIERED APPROACH

In this section, we propose a two-tiered approach to address the cluster-based HIT generation problem. We first present an overview of our approach in Section 5.1, and then discuss the top tier and the bottom tier of our approach in Sections 5.2 and 5.3, respectively.

### 5.1 Approach Overview

Similar to the approximation algorithm in Section 4, our approach first builds a graph on a set of pairs. Since our hybrid human-machine workflow typically only needs to check a small fraction of all possible pairs, the graph is very sparse, and may consist of many connected components. We classify these connected components into two types according to the cluster-size threshold $k$. *Large connected components* (LCCs) have more than $k$ vertices while *small connected component* (SCCs) have $k$ vertices or fewer.

LCCs have more vertices than can fit into a cluster-based HIT; they must be partitioned into SCCs. When partitioning, we would like to create SCCs that are highly connected since doing so increases the number of edges covered by the component, enabling more comparisons to be done in a given cluster-based HIT. The number of HITs required can also be reduced by batching multiple SCCs into a single cluster-based HIT. Different packing methods can lead to different numbers of required HITs. It is important to note that the approximation algorithm in Section 4 does not consider either of these issues that impact the number of HITs, rather it simply adds a random vertex and its corresponding edges into $SEQ$.

For example, consider the graph in Figure 5. It consists of two connected components. Suppose $k = 4$. The top one is an LCC and the bottom one is an SCC. Consider the top component containing seven vertices $\{r_1, r_2, r_3, r_4, r_5, r_6, r_7\}$. Since it is an LCC, it must be partitioned. Assume it is partitioned into three SCCs $\{r_1, r_2, r_3, r_7\}$, $\{r_3, r_4, r_5, r_6\}$, $\{r_4, r_7\}$, which can cover all of its edges. Then the graph becomes $\{r_1, r_2, r_3, r_7\}$, $\{r_3, r_4, r_5, r_6\}$, $\{r_4, r_7\}$, $\{r_8, r_9\}$.

Next, we need to pack these SCCs into cluster-based HITs. One way is to create a cluster-based HIT for each of $\{r_3, r_4, r_5, r_6\}$ and $\{r_1, r_2, r_3, r_7\}$, and then to combine $\{r_4, r_7\}$ and $\{r_8, r_9\}$ into a third cluster-based HIT. This way, we generate only three cluster-based HITs.

**Algorithm 1**: TWO-TIERED($\mathcal{P}$, $k$)

**Input**: $\mathcal{P}$ : a set of pairs of records
$k$ : a cluster-size threshold
**Output**: $H_1, H_2, \cdots, H_h$: cluster-based HITs
1 **begin**
2    Let CC denote the connected components of the graph that is built based on $\mathcal{P}$;
3    SCC = $\{cc \in CC \mid |cc| \leq k\}$; //Small Connected Components
4    LCC = $\{cc \in CC \mid |cc| > k\}$; //Large Connected Components
5    SCC $\cup$ = PARTITIONING(LCC, $k$); //Top Tier
6    $H_1, H_2, \cdots, H_h$ = PACKING(SCC, $k$); //Bottom Tier
7 **end**

**Figure 6: An overview of two-tiered approach.**

Figure 6 shows the pseudo-code for this approach. In the initial step, we build a graph based on the given set of pairs, and divide the connected components of the graph into LCCs and SCCs (Lines 2-4). We then partition each of the LCCs into small ones so that we have a collection of SCCs (Line 5). Finally, we pack all the SCCs into cluster-based HITs (Line 6).

## 5.2 LCC Partitioning (Top Tier)

We now study the top tier of our approach, that is, given an LCC, how to partition it into SCCs such that its edges can be covered by these SCCs. As discussed in Section 5.1, we aim to create SCCs that are highly connected. Based on this idea, we devise a greedy algorithm. To partition an LCC, the algorithm iteratively generates an SCC with the *highest* connectivity, and iterates until the generated SCCs cover all edges in the large one.

In each iteration step, the algorithm first initializes a small connected component scc with the vertex having the maximum degree in the LCC. Then the algorithm repeats to add a new vertex into scc that maximizes the connectivity of scc. More specifically, for each vertex $r$ ($\notin$ scc), the algorithm computes the *indegree* and the *outdegree* of $r$ w.r.t scc, where the indegree is defined as the number of edges between $r$ and the vertices in scc, and the outdegree is defined as the number of edges between $r$ and the vertices *not* in scc.

We select the vertex with the maximum indegree as that adds the most edges to scc. If there is a tie, that is, more than one vertex has the same maximum indegree, the algorithm selects the vertex with the minimum outdegree from these vertices since vertices with a larger outdegree have more connectivity with the vertices outside scc. The algorithm adds the selected vertex into scc, and updates the indegree and the outdegree of each vertex w.r.t the new scc, and repeats the above process to select another vertex. The algorithm stops adding vertices to scc when the size of scc is equal to the cluster-size threshold $k$, or when no remaining vertex connects with scc.

Figure 7 shows the pseudo-code for the top tier. Each large connected component, lcc, is partitioned into SCCs as follows. First, it creates a small connected component, scc, with the vertex that has the maximum degree in lcc (Line 5). Let conn denote a set of vertices that connect with scc (Line 6). Next, the algorithm repeatedly picks up a vertex from conn and adds it into scc until either the size of scc is $k$ or conn is empty (Lines 7-12). When picking up a vertex, it aims to maximize the connectivity of scc (Line 8). After adding a new vertex into scc (Line 9), the algorithm

**Algorithm 2**: PARTITIONING(LCC, $k$)

**Input**: LCC : a set of large connected components
$k$ : a cluster-size threshold
**Output**: SCC : a set of small connected components obtained by partitioning each large connected component in LCC
1 **begin**
2    **for** each lcc $\in$ LCC **do**
3      **while** lcc has edges **do**
4        Let $r_{max}$ be the vertex of lcc with the maximum degree;
5        scc = $\{r_{max}\}$;
6        conn = $\{r \mid$ for each edge $(r_{max}, r)$ of lcc$\}$;
7        **while** |scc| $< k$ **and** |conn| $> 0$ **do**
8          Pick up a vertex $r_{new}$ from conn with the maximum indegree w.r.t scc. (If there is a tie, pick up the one with the minimum outdegree w.r.t scc);
9          Move $r_{new}$ from conn to scc;
10          **for** each edge $(r_{new}, r)$ of lcc **do**
11            **if** $r \notin$ scc and $r \notin$ conn **then**
12              Add $r$ into conn;
13        Add scc into SCC;
14        Remove the edges of lcc that are covered by scc;
15 **end**

**Figure 7: The top tier of our approach.**

needs to update conn (Lines 10-12). Finally, the algorithm outputs scc and removes the edges of lcc that are covered by scc (Lines 13-14). If lcc still has edges, the algorithm returns to the first step and continues to generate more SSCs (Line 3).

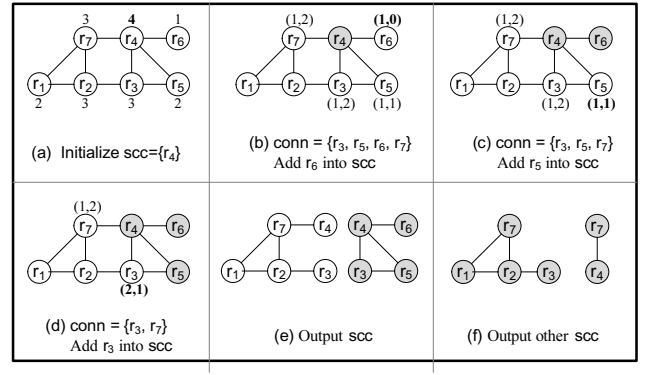

**Figure 8: An example of using the top tier of our approach to partition a large connected component in Figure 5 ($k = 4$).**

EXAMPLE 3. *For example, consider the LCC in Figure 5. To partition it into SCCs, in Figure 8(a) we first initialize scc = $\{r_4\}$ including a vertex with the maximum degree. Then we repeat to add a new vertex into scc until its size reaches the cluster-size threshold (i.e., |scc| = k) or no more vertices are available (i.e., |conn| = 0). In Figure 8(b), there are four vertices connecting with scc, thus conn = $\{r_3, r_5, r_6, r_7\}$. For each vertex, we compute its indegree and outdegree w.r.t scc, denoted by (indegree, outdegree). Since $r_3$, $r_5$, $r_6$ and $r_7$ have the same indegree while $r_6$ has the minimum outdegree, we add $r_6$ into scc. Similarly, in Figure 8(c) and (d), we respectively add $r_5$ and $r_3$ into scc.*

*At this point, if we add more vertices, |scc| will be larger than k, thus in Figure 8(e), we output scc and remove its edges from the LCC. We use the similar method to partition the remainder of the large connected component, and obtain two other SCCs in Figure 8(f). Thus, the LCC is ultimately partitioned into three SCCs: $\{r_3, r_4, r_5, r_6\}$, $\{r_1, r_2, r_3, r_7\}$ and $\{r_4, r_7\}$.*

### 5.3 SCC Packing (Bottom Tier)

We now describe the bottom tier of our approach, that is, given a set of SCCs, how to pack them into the minimum number of cluster-based HITs such that the size of each cluster-based HIT is no larger than $k$. This is a NP-Hard problem which is a variant of the one-dimensional cutting-stock problem [14]. We formulate it as an integer linear program.

Let $p = [a_1, a_2, \cdots, a_k]$ denote a *pattern* of a cluster-based HIT, where $a_j$ ($j \in [1, k]$) is the number of SCCs in the HIT that contain $j$ vertices. Since a cluster-based HIT can contain at most $k$ vertices, we say that $p$ is a feasible pattern only if $\sum_{j=1}^{k} j \cdot a_j \leq k$ holds. For example, suppose $k = 4$. $p_1 = [0, 0, 0, 1]$ is a feasible pattern since $1 \cdot 0 + 2 \cdot 0 + 3 \cdot 0 + 4 \cdot 1 = 4 \leq k$ holds. We collect all feasible patterns into a set $A = \{p_1, p_2, \cdots, p_m\}$, where $p_i = [a_{i1}, a_{i2}, \cdots, a_{ik}]$ ($i \in [1, m]$).

When packing a set of SCCs into cluster-based HITs, each HIT must correspond to a pattern in $A$. Let $x_i$ denote the number of cluster-based HITs whose pattern is $p_i$ ($i \in [1, m]$). Then the problem becomes how to minimize the total number of patterns, i.e., $\sum_{i=1}^{m} x_i$. Based on this idea, we can formulate our packing problem as the following integer linear program:

$$\min \sum_{i=1}^{m} x_i$$
$$s.t. \sum_{i=1}^{m} a_{ij} x_i \geq c_j, \quad \forall j \in [1, k]$$
$$x_i \geq 0, \quad \text{integer}$$

where $c_j$ is the total number of the small connected components that contain $j$ vertices.

For example, given a set of SCCs, $\{r_3, r_4, r_5, r_6\}$, $\{r_1, r_2, r_3, r_7\}$, $\{r_4, r_7\}$ and $\{r_8, r_9\}$, we have $c_1 = 0$, $c_2 = 2$, $c_3 = 0$ and $c_4 = 2$. To pack them into cluster-based HITs ($k = 4$), we first generate all feasible patterns, i.e. $A = \{p_1 = [0, 0, 0, 1], p_2 = [0, 2, 0, 0], p_3 = [0, 1, 0, 0]\}$. (Note that since $c_1 = 0$ and $c_3 = 0$, we omit the feasible patterns in $A$ whose first or third dimension contains non-zero values.)

Next we need to decide the number of cluster-based HITs corresponding to each feasible pattern, i.e. $x_1$, $x_2$ and $x_3$. One possible solution is $x_1 = 2$, $x_2 = 0$ and $x_3 = 2$, which needs $\sum_{i=1}^{3} x_i = 4$ cluster-based HITs. We can easily verify that the solution satisfies the constraint condition, i.e. $\sum_{i=1}^{3} a_{ij} x_i \geq c_j$ ($\forall j \in [1, 4]$). For example, when $j = 2$, we have $\sum_{i=1}^{3} a_{i2} x_i = 0 \cdot 2 + 2 \cdot 0 + 1 \cdot 2 \geq c_2 = 2$. However, this solution is not optimal since there is another solution $x_1 = 2$, $x_2 = 1$ and $x_3 = 0$ that also satisfies the constraint condition but needs only $\sum_{i=1}^{3} x_i = 3$ cluster-based HITs.

The above integer linear program can be solved by using column generation and branch-and-bound [25]. The technique is very efficient as it does not need to generate all feasible patterns at the beginning. Instead, it starts with a few patterns and generates more patterns as needed. At each iteration, a branch-and-bound tree is built to search for the optimal integer solution.

## 6. BACK OF THE ENVELOPE ANALYSIS

In this section, we compare pair-based HITs with cluster-based HITs analytically in terms of the number of comparisons they require workers to perform. For a pair-based HIT the number of comparisons required is simply the number of pairs that have been batched into the HIT. This is because each pair in the HIT is treated separately. For cluster-based HITs the story is more involved.

Consider a cluster-based HIT with $n$ records. One may be tempted to think that $\frac{n \cdot (n-1)}{2}$ comparisons would be required. However, in reality, the number of comparisons also depends on the way that a person does the HIT and the number of distinct entities represented in the HIT. Suppose the cluster-based HIT contains $m$ distinct entities, denoted by $e_1, e_2, \cdots, e_m$, where $e_i$ ($i \in [1, m]$) represents the set of records in the HIT that refer to the $i$-th entity. Obviously, $\sum_{i=1}^{m} |e_i| = n$.

Assume a person does the cluster-based HIT as follows. First, she picks up a record from an entity, e.g. $e_1$. Then she compares the record with the other $n - 1$ records. After $n - 1$ comparisons, she can identify all the records in the cluster-based HIT that refer to $e_1$. Next she selects a record from another entity, e.g. $e_2$. Note that she does not need to compare it with the records in $e_1$ since those records correspond to the first entity and cannot refer to the second (different) entity. She can then identify all the records that refer to $e_2$ with $n - 1 - |e_1|$ comparisons. Iteratively, when selecting a record from $e_i$, she only needs to compare with $n - 1 - \sum_{j=1}^{i-1} |e_j|$ other records. By summing the number of comparisons in each iteration step, we can obtain the total number of comparisons required to complete the cluster-based HIT, i.e.,

$$\sum_{i=1}^{m} \left( n - 1 - \sum_{j=1}^{i-1} |e_j| \right). \quad (1)$$

Based on this equation, we have the following two observations.

First, the value of Equation 1 decreases as $|e_j|$ ($j \in [1, m]$) increases. That is, a cluster-based HIT requires fewer comparisons when it contains more matches (i.e., duplicates). To illustrate this observation, consider two extreme cases. One is no duplicate record exists in a cluster-based HIT. For this case, there are $n$ entities in the cluster-based HIT and each entity contains only one record, thus the number of comparisons becomes $\frac{n \cdot (n-1)}{2}$. The other case is all records in a cluster-based HIT are duplicate. For this case, there is only one entity in the cluster-based HIT and the entity contains $n$ records and the number of comparisons required is $n - 1$.

The second observation is that the value of Equation 1 differs in the sequence of the identified entities. For ease of presentation, we modify Equation 1 to the following equivalent equation.

$$(n - 1) \cdot m - \sum_{i=1}^{m-1} (m - i) \cdot |e_i|. \quad (2)$$

The first part of the equation, i.e. $(n - 1) \cdot m$, is a constant value w.r.t the sequence of the entities, while the second part, i.e. $\sum_{i=1}^{m-1}(m-i) \cdot |e_i|$, is a weighted sum of $|e_i|$ ($i \in [1, m-1]$). Since the weight, i.e., $(m-i)$, decreases with increasing $i$, the best way to identify entities is in increasing order of $|e_i|$, which can result in the minimum number of comparisons. On the other hand, the worst way to identify entities is in decreasing order of $|e_i|$.

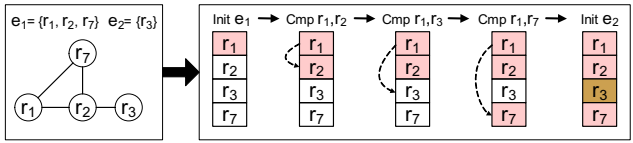

**Figure 9: An illustration of computing the number of comparisons for a cluster-based HIT.**

Example 4 shows how to use the above method to compute the number of comparisons for a cluster-based HIT.

EXAMPLE 4. *In Figure 9, consider a cluster-based HIT with $n = 4$ records, $\{r_1, r_2, r_3, r_7\}$. From Table 1, we can see that $r_1$, $r_2$, $r_7$ refer to the same entity, thus $e_1 = \{r_1, r_2, r_7\}$ and $e_2 = \{r_3\}$. Figure 9 shows the way a human worker does the HIT. First, the human initializes an entity $e_1$ by selecting a record $r_1$. Then she compares $r_1$ with the other $n - 1 = 3$ records. Since $r_2$ and $r_7$ refer to the same entity as $r_1$, she adds them into $e_1$ by painting them the same color as $r_1$. We can see that after three comparisons, all of the records that refer to $e_1$ have been identified. Next, the human worker initializes another entity $e_2$ by selecting record $r_3$. Since no record is left in the cluster-based HIT, i.e. $n - 1 - |e_1| = 0$, there is no need to compare $r_3$ with any record. Therefore, the cluster-based HIT requires only three comparisons in total. One interesting observation is the human worker actually checks four pairs of records $(r_1, r_2)$, $(r_1, r_7)$, $(r_2, r_3)$ and $(r_2, r_7)$ using only three comparisons. A pair-based HIT would require four comparisons. This shows that cluster-based HITs can require fewer comparisons than pair-based HITs.*

## 7. EXPERIMENTAL RESULTS

We conducted extensive experiments to evaluate our approaches. We address three issues here. First, we examine the effectiveness of our two-tiered HIT generation approach in reducing the number of HITs required for Entity Resolution on real data sets. Second, we compare the quality of the results produced by our hybrid human-machine approach with that produced by two machine-based approaches. Finally, we compare pair-based and cluster-based HITs in terms of both answer quality and latency.

### 7.1 Experimental Setup

**Datasets**: We used two real datasets to evaluate our method.

*Restaurant*[2] is a data set consisting of 858 (non-identical) restaurant records. It has $\frac{858 \cdot (858-1)}{2} = 367,653$ pairs of records in total, among which 106 pairs refer to the same entity. Each restaurant record has four attributes, [*name, address, city, type*]. An example record is: ["oceana", "55 e. 54th st.", "new york", "seafood"].

[2] http://www.cs.utexas.edu/users/ml/riddle/restaurant.tar.gz

**Table 2: Likelihood-threshold selection.**

(a) *Restaurant* Dataset

| Threshold | Total #Pair | Matches | Recall |
|---|---|---|---|
| 0.5 | 161 | 83 | 78.3% |
| 0.4 | 755 | 99 | 93.4% |
| 0.3 | 4,788 | 105 | 99.1% |
| 0.2 | 23,944 | 106 | 100% |
| 0.1 | 83,117 | 106 | 100% |
| 0 | 367,653 | 106 | 100% |

(b) *Product* Dataset

| Threshold | Total #Pair | Matches | Recall |
|---|---|---|---|
| 0.5 | 637 | 335 | 30.5% |
| 0.4 | 1,427 | 571 | 52.1% |
| 0.3 | 3,154 | 805 | 73.4% |
| 0.2 | 8,315 | 1,011 | 92.2% |
| 0.1 | 37,641 | 1,090 | 99.4% |
| 0 | 1,180,452 | 1,097 | 100% |

*Product*[3] is a product data set integrated from two different sources. There are 1081 records coming from the abt website and 1092 records coming from the buy website. The data set has $1081 * 1092 = 1,180,452$ pairs of records, among which 1,097 pairs refer to the same entity. Each product record has two attributes, [*name, price*]. An example record is: ["Apple 8GB Black 2nd Generation iPod Touch - MB528LLA", "$229.00"].

The two datasets were preprocessed by replacing non-alphanumeric characters with white spaces, and letters with their lowercases.

**Machine-based Technique**: Our hybrid human-machine workflow needs a machine-based technique to compute a likelihood for each pair of records. In our experiment, a simple similarity-based technique, called simjoin, was adopted to achieve this goal. We first generated a token set for each record, which consisted of the tokens from all attribute values. Then for each pair of records, we took the Jaccard similarity between their corresponding token sets as their likelihood. Since our workflow only crowdsources the pairs whose likelihood is above a threshold, Table 2 shows the effect of different selections of thresholds on the two datasets. For example, for the *Restaurant* dataset, a threshold of 0.5 retains 161 pairs. In reality 83 of these pairs refer to the same entity. The recall is $\frac{83}{106} = 78.3\%$, which means that 78.3% matching pairs out of the total 106 matching pairs in the data set pass the threshold. From Table 2, we can conclude that a hybrid human-machine workflow can utilize machine-based techniques to significantly reduce the number of the pairs with a little loss of recall. For instance, on the *Product* dataset, when the threshold is 0.2, we can achieve up to 92.2% recall by having people examine only 8,315 pairs of records, which is over two orders of magnitude fewer than the total number of pairs (1,180,452).

**AMT**: We use Amazon Mechanical Turk (AMT) to evaluate our hybrid human-machine workflow. AMT is a widely used crowdsourcing marketplace. We paid workers $0.02 for completing each HIT and paid AMT $0.005 for publishing each HIT. We ran over 8000 HITs and spent about $600 on AMT to evaluate our methods. All the HITs were published between 1800 and 2400 PST. We ran each experiment three

[3] http://dbs.uni-leipzig.de/file/Abt-Buy.zip

times at the same time of day during the course of three days, and report the average performance. In addition, we used two ways to improve the result quality. (1) *Assignment*: AMT allows us to replicate one HIT into multiple assignments, and guarantees that each assignment can be done by a different worker. In our experiment, each HIT was replicated into 3 assignments. That is, we obtained the results of a HIT from three different workers and made our final decision based on a combination of the three results (see Section 7.3). (2) *Qualification Test*: We found that some workers may do our HITs maliciously. In order to prevent this, AMT supports qualification tests for workers, and only those who successfully pass the test can do our HITs. In our experiment, the qualification test consists of three pairs of records. For each one, a worker needs to decide whether or not they match. Workers must get all three pairs correct to pass the qualification test.

## 7.2 Cluster-based HIT Generation

In this section, we evaluate the two-tiered approach for cluster-based HIT generation. We compare with the following baseline algorithms:

**Random**: The algorithm generates cluster-based HITs by randomly selecting records from a set of pairs of records, $\mathcal{P}$. To generate a cluster-based HIT, $H$, it repeatedly selects a pair of records from $\mathcal{P}$ and merges the two records into $H$. When $|H| = k$, it outputs $H$, and removes the pairs from $\mathcal{P}$. If $\mathcal{P}$ still has pairs, the algorithm will repeat the above process to generate new cluster-based HITs; otherwise, the algorithm terminates.

**BFS-based**: The algorithm first builds a graph on a given set of pairs of records, and then generates cluster-based HITs according to the breadth-first-search (BFS) of the graph. To generate a cluster-based HIT, $H$, the algorithm traverses the graph using BFS, and adds the vertices (i.e. records) into $H$ in the traversal order. When $|H| = k$, it outputs $H$, and removes the edges that can be covered by $H$ from the graph. If the graph still has edges, the algorithm will repeat the above process to generate new cluster-based HITs; otherwise, the algorithm terminates.

**DFS-based**: Similar to BFS-based but traverses the graph using depth-first-search (DFS).

**Approximation**: The k-clique approximation algorithm described in Section 4.

We first compare the two-tiered approach with the baseline algorithms for various likelihood thresholds. We varied the likelihood threshold from 0.5 to 0.1 on the *Restaurant* and *Product* datasets, and used the different approaches to generate cluster-based HITs. Figure 10 shows the number of generated cluster-based HITs ($k = 10$) as the threshold is varied. We can see that the two-tiered approach generated the fewest cluster-based HITs, with the differences being greater for smaller thresholds. Note that in order to achieve a high recall, we need to select a smaller threshold (Table 2).

In terms of the baseline algorithms, we have the following observations. First, the BFS-based algorithm was the best baseline algorithm. This is because the BFS traversal of the graph can generate cluster-based HITs with highly connected vertices. Second, the approximation algorithm did not perform well on the real datasets. For example, on the *Restaurant* dataset, when the threshold is 0.1, it performed worst. Third, the naive random algorithm generated many more cluster-based HITs than the two-tiered approach. For example, on the *Product* dataset, the random algorithm generated 6422 HITs with threshold 0.1, while the two-tiered approach generated only 2033 HITs.

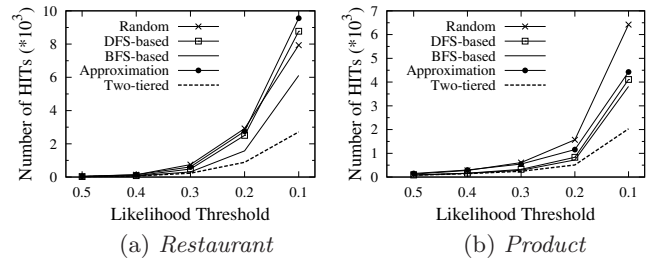

**Figure 10: Comparison of the number of cluster-based HITs for various likelihood thresholds (cluster-size=10).**

Next we compare two-tiered approach with the baseline algorithms for various cluster-size thresholds. We varied the threshold from 5 to 20 on the *Restaurant* and *Product* datasets, and compared the number of cluster-based HITs generated by different approaches. Figure 11 shows the results with likelihood threshold = 0.1. We can see that for all cluster-size thresholds tested, our two-tiered approach generated the minimum number of cluster-based HITs. For example, on the *Restaurant* dataset the two-tiered approach generated 1.9 to 2.3 times fewer cluster-based HITs than the best baseline algorithm.

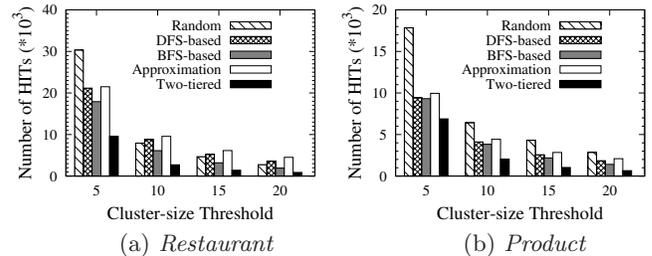

**Figure 11: Comparison of the number of cluster-based HITs for various cluster-size thresholds (likelihood threshold=0.1).**

## 7.3 Entity-Resolution Techniques

In this section, we compare the hybrid human-machine workflow with existing entity resolution techniques. We use two metrics to evaluate the result quality: (1) *precision* is the percentage of correctly identified matching pairs out of all pairs identified as matches; (2) *recall* is the percentage of correctly identified matching pairs out of all matching pairs in the dataset. As more matches are identified, recall increases while precision potentially decreases.

We show our results as precision-recall curves [4], generated as follows. We assume the result of an entity-resolution technique is a ranked list of pairs of records, where the list is sorted based on the decreasing order of the likelihood that a pair of records match. In the list, the first $n$ pairs are identified as matching pairs. To plot the precision-recall curve, we vary $n$ and plot the precision vs. the recall.

We implemented the following entity-resolution techniques.

**simjoin:** This is the machine-based technique used by our hybrid human-machine workflow (see Section 7.1).

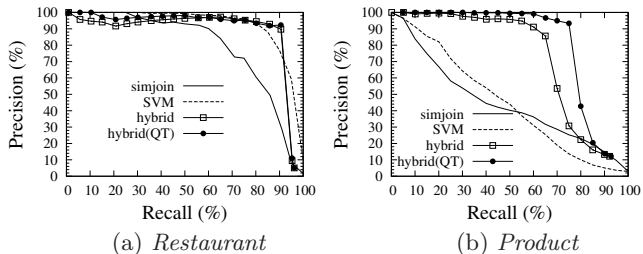

(a) *Restaurant*  (b) *Product*

Figure 12: Comparing hybrid human-machine workflow with existing machine-based techniques.

**SVM:** This is a state-of-the-art learning-based technique. First, we computed a feature vector for each pair of records. For the *Restaurant* dataset, we chose two similarity functions, *edit distance* and *cosine similarity*, adopted by [18], and computed their values on four attributes to obtain a 8-dimensional feature vector; For the *Product* dataset, we chose the same two similarity functions and computed their values on the *Name* attribute to obtain a 2-dimensional feature vector. Next we trained a classifier on 500 pairs that were randomly selected from the pairs whose Jaccard similarities were above 0.1 (Note that the training pairs were sampled 10 times, and we report the average performance here). Finally, SVM returned a ranked list of the remaining pairs sorted based on the likelihood given by the classifier [4].

**hybrid:** Our hybrid human-machine workflow first uses simjoin to obtain a set of pairs based on a specified threshold, and then verifies these pairs by using the cluster-based HITs generated by the two-tiered approach with $k = 10$. For the *Restaurant* dataset, simjoin returned 2004 pairs (102 matching pairs, 96.2% recall) based on the specified threshold 0.35, and the two-tiered approach generated 112 cluster-based HITs. On the *Product* dataset, simjoin returned 8315 pairs (1,011 duplicate pairs, 92.2% recall) based on the specified threshold 0.2, and the two-tiered approach generated 508 cluster-based HITs. We posted these HITs on AMT and replicated each HIT into three assignments. Thus, we spent $112 * 3 * 0.025 = \$8.4$ on the *Restaurant* dataset, and $508 * 3 * 0.025 = \$38.1$ on the *Product* dataset. Finally, the hybrid human-machine workflow returned a ranked list of the pairs sorted based on the results of the crowd.

One detail we need to mention is the way we combined the answers from the three different assignments for each HIT. A simple technique would be to average the three responses for each HIT, but this approach is susceptible to spammers. Instead we adopted the EM-based algorithm [9], which has been shown to be effective in previous work [16,19].

We first compare hybrid with simjoin and SVM on the *Restaurant* and *Product* datasets. Figure 12 shows the results. In the figure, hybrid and hybrid(QT) respectively denote the hybrid workflow with and without a qualification test. For the *Restaurant* dataset, we can see that hybrid and hybrid(QT) achieve the same quality as SVM. This indicates that the hybrid human-machine workflow based on a simple non-learning based technique (i.e. simjoin) can have a comparable performance to a sophisticated learning based technique (i.e. SVM). On the *Product* dataset, we can see that hybrid and hybrid(QT) achieved significantly better quality than simjoin and SVM. This indicates that for datasets for which the machine-based techniques were unable to perform well, a hybrid human-machine workflow still can achieve very high quality.

Note that to further improve the recall of the hybrid workflow, we can specify a smaller likelihood threshold thereby asking the crowd to perform more HITs. For example, in Figure 12(b), hybrid and hybrid(QT) can achieve at most 92.2% recall. In contrast, as shown in Table 2(b), our hybrid workflow, if used with a likelihood threshold of 0.1, could achieve up to 99.4% recall at the cost of crowdsourcing 37,641 pairs.

Next we compare hybrid with hybrid(QT). The results in Figure 12 show that adding a qualification test can in fact help to improve the result quality. There are mainly two reasons for this. First, a qualification test can weed out spammers since they are very likely to fail the test. Second, is that a qualification test can force workers to read our instructions more carefully. However, while the qualification test can improve quality, this improvement may come at a steep cost in terms of latency. For the *Restaurant* dataset, hybrid and hybrid(QT) required 1.3 hours and 1.6 hours respectively to complete 112 HITs; on the *Product* dataset, hybrid and hybrid(QT) required 4.5 hours and 19.9 hours respectively to complete 508 HITs.

### 7.4 Pair-based vs. Cluster-based HITs

Having shown the benefits of hybrid Entity Resolution compared to both machine-based and human-based methods, we now turn to examining the relative performance of pair-based vs. cluster-based HITs. As described in Section 6, the benefit of the cluster based approach in terms of number of comparisons depends on the number of matching pairs in the data set. Thus, in this comparison, in addition to the *Product* data set used previously, we also use an additional dataset we created called *Product+Dup* that has more matching pairs than the datasets used in the previous experiences.

We created the *Product+Dup* by randomly selecting 100 records from the *Product* dataset, and then adding $x$ matching records for each base record, where $x$ is an integer random variable uniformly distributed on $[0, 9]$. Matches were generated by randomly swapping two tokens in the base record. *Product+Dup* has 157,641 pairs of records, among which 1713 pairs are matches.

We generated pair-based and cluster-based HITs using a likelihood threshold of 0.2. We set the cluster size ($k = 10$), which we denote by $C_{10}$. In order to keep cost constant across the two methods, we created pair-based HITs that contained enough pairs so that the number of HITs generated for both methods was the same. For the *Product* dataset, there were 8315 pairs that needed to be crowdsourced, resulting in 508 cluster-based HITs. In order to generate the same number of pair-based HITs, we needed to generate pair-based HITs containing $\frac{8315}{508} = 16$ pairs on average, denoted by $P_{16}$. Similarly, for the *Product+Dup* dataset, there were 3401 pairs that needed to be crowdsourced, resulting in 120 cluster-based HITs, and 120 pair-based HITs, containing ($\frac{3401}{120} = 28$) pairs on average, denoted by $P_{28}$.

We first compare the median completion time per assignment between pair-based HITs and cluster-based HITs. Figure 13 shows the results. *QT* represents the experimental results with a qualification test. We can see that the time to complete a single HIT was lower for the cluster-based HITs than for the pair-based HITs in these experiments. For the *Product* dataset, Figure 13(a), a cluster-based HIT consumed about 15% less time than a pair-based HIT. The difference was more dramatic for the *Product+Dup* dataset, which contains more matching pairs.

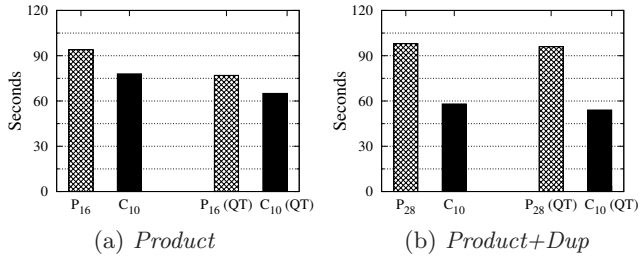

(a) *Product*  (b) *Product+Dup*

**Figure 13: Comparison of median completion time per assignment between a pair-based HIT and a cluster-based HIT.**

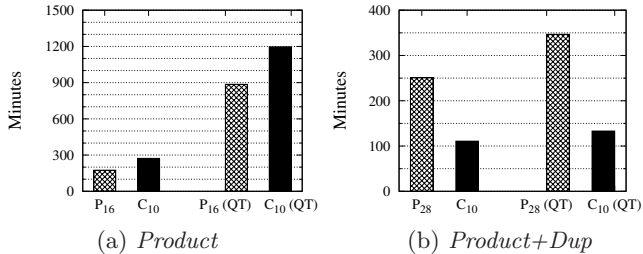

(a) *Product*  (b) *Product+Dup*

**Figure 14: Comparison of the time of completing all pair-based HITs and all cluster-based HITs.**

However, the results are somewhat different when considering the total time taken to receive all of the results (rather than a single HIT). These results are shown in Figure 14. Surprisingly, for the *Product* dataset, pairs-based HITs were completed earlier. We find the pair-based HITs attracted more workers. This may be due to the unfamiliar interface of cluster-based HITs that makes workers feel that it is harder to complete. Since we did these experiments on AMT, we did not have specifically trained workers who were familar with the cluster-based interface. For the *Product+Dup* dataset, however, the efficiency of cluster-based HITs in the presence of more matches led to advantages in overall completion time as well. Recall that pair-based HITs containing on average 28 pairs were required to produce the same number of HITs as the cluster-based approach, compared to only 16 for the *Product* dataset. Since we kept the price per HIT constant, fewer workers were attracted to perform the pair-based HITs in this case.

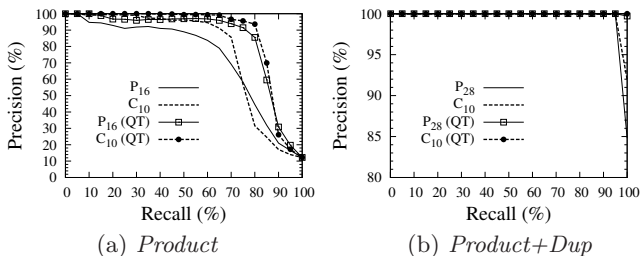

(a) *Product*  (b) *Product+Dup*

**Figure 15: Comparison of the quality of pair-based HITs and cluster-based HITs.**

Finally, Figure 15 shows that the result quality for pair-based HITs and cluster-based HITs was similar in these experiments.

## 8. RELATED WORK

Entity resolution is a critical task for data integration and cleaning. It has been studied extensively for several decades (see [11] for a survey). Some existing work has investigated how to benefit from human interaction. Sarawagi et al. [24] proposed an entity-resolution method using active learning, which allows a user to label a training set interactively. They have shown the method can significantly reduce the size of the training set needed to achieve high accuracy. Arasu et al. [1] proposed another active-learning approach which is more scalable to large datasets and is able to give probabilistic guarantees on the quality of the results. Jeffery et al. [17] studied user feedback in pay-as-you-go data integration systems. In such systems, there exist some candidate duplicate pairs of elements (e.g. attribute names or values) that require user feedback for verification. They proposed a decision-theoretic approach to determine the order in which these pairs should be verified. McCann et al. [21] studied schema matching in online communities. They generated different types of questions to ask community members, and derived schema-matching results from the answers of these questions.

Recently, crowdsourcing has attracted significant attention in both the industrial and academic communities (see [10] for a recent survey). Recent projects in the database community aim to embed crowdsourcing into database query processing. Franklin et al. [12,13] extended relational database query language SQL to CrowdSQL by enabling crowd-based operators, and built CrowdDB, a relational query processing system based on CrowdSQL. Marcus et al. [19] integrated SQL with crowd-based user defined functions (UDFs), and proposed Qurk, a declarative workflow management system. Parameswaran et al. [22] presented Deco, a database system for declarative crowdsourcing. In addition, there are many hybrid human-machine systems being developed outside of the database community. CrowdSearch [27] is an image searching system, that combines automated image search with real-time human validations of search results. Solyent [3] is a word processor that utilizes crowd workers to shorten, proofread, and edit documents. Although there are many studies in crowdsourcing, to the best of our knowledge, no existing work has explored how to improve entity resolution using hybrid human-machine techniques combining a generic microtask crowdsourcing platform with machine-based techniques.

There are also some studies on blocking which consider partitioning of a table of records to maximize matching record pairs co-occurring in given partitions [7]. Although our cluster-based HIT generation problem is a form of blocking, it differs from the typical blocking problem. Firstly, the fact that our block size is constrained by what people can do is different than what determines block size typically (which is that beyond a certain point, increasing the block size does not reduce the complexity). Secondly, since the financial cost of human comparisons is driven by the number of tasks (i.e., blocks), our goal is to minimize number of blocks of the given size, which is a different objective than that of previous work.

## 9. CONCLUSION AND FUTURE WORK

In this paper we have studied the problem of crowdsourcing entity resolution. We described how machine-only approaches often fall short on quality, while brute force people-only approaches are too slow and expensive. Thus, we proposed a hybrid human-machine workflow to address this problem. In the context of this hybrid approach, we investigated pair-based and cluster-based HIT generation. In particular, we formulated the cluster-based HIT generation problem, and showed that it is NP-Hard. We then devised a heuristic two-tiered approach to solve this problem. We described this method, and presented the results of extensive experiments on real data sets using the AMT platform. The experiments show that our hybrid approach achieves both good efficiency and high accuracy compared to machine-only or human-only alternatives. In particular, the results indicated that (1) the two-tiered approach generated fewer cluster-based HITs than existing algorithms; (2) hybrid human-machine workflow significantly reduced the number of HITs compared to human-based techniques, and achieved higher quality than the state-of-the-art machine-based techniques; and (3) the cluster-based HITs can provide lower latency than a pair-based approach, particularly in the presence of many matching records, but that the simplicity of the pair-based interface seemed to be appealing to AMT workers.

Our work represents an initial approach towards hybrid human-machine entity resolution. There are many further research directions to explore. First of all, we were surprised that some AMT workers preferred to do the relatively large pair-based tasks over the much smaller cluster-based tasks, even though the price paid for them was identical. We believe that this could be due in part to worker's lack of familiarity with the cluster-based interface, which if true, raises the possibility that different approaches could have very different performance when applied to experienced vs. novice crowds. There are also many user interface improvements that could be made to both the pair-based and cluster-based interfaces, which could have dramatic effects on cost, quality and latency.

Another issue to be addressed is that of scaling to much larger datasets. Our approach utilizes machine-based techniques to remove dissimilar pairs. However, for a data set with millions of records, there will be a large number of remaining pairs need to be verified by the crowd. Therefore, we need to explore how to make a better use of machine-based techniques to further offload relatively expensive crowd resources. A related issue is the development of a budget-based approach to hybrid entity resolution. Users may wish to trade off cost, quality and latency, and the development of tools and algorithms to facilitate such tradeoffs is a deep research challenge. Finally, we would like to extend these techniques to take privacy into consideration. Sometimes, the data to be integrated is confidential, so other techniques will be required to allow crowds to process the data.

**Acknowledgements.** This research is supported in part by NSF CISE Expeditions award CCF-1139158, gifts from Amazon Web Services, Google, SAP, Blue Goji, Cisco, Cloudera, Ericsson, General Electric, Hewlett Packard, Huawei, Intel, MarkLogic, Microsoft, NetApp, Oracle, Quanta, Splunk, VMware and by DARPA (contract #FA8650-11-C-7136).

Jianan Wang and Jianhua Feng were partly supported by the National Natural Science Foundation of China under Grant No. 61003004, the National Grand Fundamental Research 973 Program of China under Grant No. 2011CB302206, a project of Tsinghua University under Grant No. 20111081073, and the "NExT Research Center" funded by MDA, Singapore, under the Grant No. WBS:R-252-300-001-490.